\documentstyle[psfig]{cici}

\def\simg{\mathrel{%
      \rlap{\raise 0.511ex \hbox{$>$}}{\lower 0.511ex \hbox{$\sim$}}}}
\def\siml{\mathrel{%
      \rlap{\raise 0.511ex \hbox{$<$}}{\lower 0.511ex \hbox{$\sim$}}}}
\def\etal{et al$.$ } \def\eg{e$.$g$.$ } \def\ie{i$.$e$.$ } 
\def\deg{^{\rm o}} \def\cm3{\rm \,cm^{-3}} \def\ergsr{\rm \,erg\;sr^{-1}} 

\def\reference{\bibitem}

\begin{document}

\title[ GRB Afterglows 050709 and 050724 ]
       {The Energetics and Environment of the Short-GRB Afterglows 050709 and 050724}

\author[A. Panaitescu]{A. Panaitescu \\
        Space Science and Applications, MS D466, Los Alamos National Laboratory, Los Alamos, NM 87545 }

\maketitle

\begin{abstract}
\begin{small}
 We use the available radio, optical and X-ray measurements for the afterglows of the short 
 bursts 050709 and 050724 to constrain the blast-wave energy, its collimation and the density 
 of the circumburst medium. For GRB 050709 (whose duration was 0.07 s), we identify two kinds 
 of models: $i)$ a high-density solution, where the ejecta are collimated in a jet of half-angle 
 $\theta_{jet} > 6\deg$ and interact with a medium of particle density $10^{-4} \cm3 < n < 0.1 
 \cm3$, $ii)$ a low-density solution with $\theta_{jet} > 2\deg$ and $n < 10^{-5} \cm3$.
 These density ranges are compatible with those expected in the vicinity of the host galaxy and 
 in the intergalactic medium, lending support to the hypothesis that the progenitor of GRB 050907 
 is a NS-NS or NS-BH merger. For GRB 050724 (which last 3 s), we obtain $0.1 \cm3 < n < 10^3 \cm3$ 
 and $\theta_{jet} > 8\deg$. The range of allowed densities shows that this burst occurred in 
 the intragalactic environment. The dynamical parameters of the high-density model for the GRB 
 afterglow 050709 are similar to those for 050724. If these parameters are representative for 
 short-GRB outflows, then these jets are less collimated and have a lower kinetic energy than 
 those of long bursts, which suggests that GRB jets are not magnetically collimated and are 
 powered by the gravitational energy of the torus. Evidently, the analysis of more short-GRB 
 afterglows is required for a more robust conclusion.
\end{small}
\end{abstract}


\section{Introduction}

 The distribution of the logarithm of Gamma-Ray Burst (GRB) durations is bimodal (\eg Kouveliotou
\etal 1993), with a gap at about 2 seconds: a third of all BATSE bursts last between 0.01 s 
and 2 s (the short GRBS), while the 25 keV--1 MeV of the remaining two thirds extended over 
2 to 500 s (the long GRBs). Until several months ago, space-based observatories have provided 
arcminute localizations only for long bursts, enabling their optical follow-up and leading to 
the spectroscopic identification of type Ib/c supernovae associated with at least four bursts 
(GRBs 980425 -- Galama \etal 1998, 011211 -- Garnavich \etal 2003, 030329 -- Stanek \etal 2003,
031203 -- Malesani \etal 2004). The optical emission of the afterglows of several other long 
bursts exhibited a plateau at 10--30 days after the burst, suggesting a supernova contribution 
in these cases as well.

 In the last half-year, the HETE-II and Swift missions have localized accurately short GRBs
at an average rate of one per month: four bursts (050509B, 050709, 050906, 051105) last
less than 0.1 s and two other (050724, 050813) had a duration of about 1 second\footnote{
The soft tail emission of GRB 050709, extending over 150 s, and the soft second peak of GRB 
050724, occurring after 1 s, may have not been detectable to BATSE and, hence, these bursts 
can be classified as short}. Optical searches for the accompanying afterglow emission have 
yielded a few pieces of evidence in favour of a compact binary merger 
(NS-NS or NS-BH) origin for these short bursts. 

 These pieces of evidence are based on our belief that the ultra-relativistic outflows 
required to produce a GRB results from either the merger of two compact objects, one of which 
is a NS (\eg Lattimer \& Schramm 1976, Blinnikov \etal 1984) or in the collapse of a massive 
star's core (Woosley 1993, Paczy\'nski 1998). For the former model, the dynamical timescale is 
of order 10 ms, while for the latter, the gas fall-back time on the BH should last more than a 
few seconds. There are other processes which can determine the burst duration (viscous 
accretion time of the debris disk, jet penetration through its surroundings, dissipation
of outflow's energy), still the above timescales naturally lead to the association of short 
GRBs with compact mergers and of long bursts with massive stars.

 Consequently, any observational evidence that a short GRBs is not associated with a massive star:  
$i)$ a low star-formation rate for the GRB host galaxy (GRB 050509B -- Bloom \etal 2005, 
      GRB 050724 -- Berger \etal 2005, GRB 050813 -- Prochaska \etal 2005), 
$ii)$ a host galaxy with an evolved stellar population (GRB 050724 -- Gorosabel \etal 2005), 
$iii)$ the lack of a bright/blue star-forming region at the location of the afterglow 
     (GRB 050709 -- Fox \etal 2005, Hjorth \etal 2005), or 
$iv)$ the lack of an accompanying supernova (GRB 050509B -- Bloom \etal 2005, GRB 050709 -- 
     Fox \etal 2005)\footnotetext{But note the lack of a supernova associated with the long 
     GRB 040701 ($z=0.21$) down to $R=27.5$ (Soderberg \etal 2005). The redshift of this burst 
     and depth of the supernova search are comparable to those for the short GRB 050709 ($z=0.16$).},
 are (indirect) arguments in favour of the binary merger model.

 A direct proof for merging binaries as the progenitors of short GRBs would be the systematic 
lack down to $R \simg 26$ (\ie below the magnitude of the dimmer hosts of long GRBs) of a galaxy 
superposed on bursts localized with (sub-)arcsecond accuracy through their optical and X-ray 
afterglows, as the natal kick of neutron stars ($\sim 200-1,000\, {\rm km\,s^{-1}}$) can be 
sufficiently large to allow them to escape the host galaxy and travel tens to hundreds of kpcs 
during their merging times ($\sim 0.1-1$ Gyr). One difficulty here is that the afterglows of 
short bursts occurring in the intergalactic medium could be too dim to detect and localize
accurately, thus there could be an observational bias against those short bursts which escape
their hosts.

 In this work, we investigate another possible test for the origin of short burst in binary 
mergers, by modelling the multiwavelength emission of the GRB afterglows 050709 and 050724. 
Afterglow modelling constrains some dynamical parameters of the GRB outflow which can be related 
to the properties of the burst progenitor: $i)$ circumburst medium density, $ii)$ blast-wave 
kinetic energy and $iii)$ outflow collimation. 

 $i)$. If short bursts do indeed arise from binary mergers then the circumburst density resulting 
from afterglow fitting should sometimes be characteristic for the intergalactic medium (\ie of 
order $10^{-6} \cm3$). However, the inferred circumburst medium density could be characteristic 
for the interstellar medium ($0.1-1 \cm3$) or for a galactic halo (around $10^{-3} \cm3$) if the 
compact binary had short merging time or did not escape from the host. Therefore, in the case
of a short-GRB afterglow for which a spatially coincident galaxy is not observed, the circumburst
density obtained through afterglow modelling will distinguish between a truly escaped binary
interacting with the intergalactic medium and a merger occurring within a high-redshift, 
undetectable host. 

 $ii)$. If the gravitational binding energy of the torus is the source of the GRB outflow kinetic energy 
then the outflows of short GRBs could be 10 times less energetic than that of long bursts because 
the mass of the torus resulting in a binary merger ($\sim 0.1\, M_\odot$ -- Ruffert \& Janka 1998) 
is a factor 10 lower than that resulting from the collapse of a WR star's core ($\sim 1\, M_\odot$ 
-- Fryer \& Woosley 1998). Under the assumptions that the redshifts and efficiencies (ratio of 
burst $\gamma$-ray output to the outflow kinetic energy) of the short and long BATSE bursts are
similar, this energy ratio is consistent with that the fluence of short GRBs is, on average, 30 
times smaller than that of long bursts. However, as noted by M\'esz\'aros, Rees and Wijers (1999), 
the relativistic outflows resulting from binary mergers could be more energetic than those produced 
by collapsars if the magnetic field in the torus threads the BH and tap its spin energy. 

$iii)$. The constraints on the opening of short-GRB outflows may also provide some insight on the mechanism 
which collimates GRB jets. If jets are magnetically collimated then the stronger fields expected 
in compact mergers could produce narrower jets than those resulting from collapsars. However the
opposite is possible if the collimation of binary-merger short-GRB jets is due to their interaction 
with the baryonic wind from the disk (Rosswog \& Ramirez-Ruiz 2003) while that of massive-star
long-GRB jets is caused by their penetration through the core and stellar envelope (Aloy \etal 2000,
MacFadyen, Woosley \& Heger 2001).

\section{The Afterglow Model}
\label{aglow}

 The afterglow model employed here is that of a relativistic blast-wave decelerated by the 
sweeping-up of the circumburst medium. Its formalism is that described in Panaitescu \& Kumar 
(2001). In calculating the dynamics of the decelerating GRB remnant, we consider that the 
outflow is uniform, in the sense that the ejecta kinetic energy per solid angle ($E$) is the 
same in any direction. The scarcity of the available data does not warrant the use of a 
structure outflow. Radiative losses are taken into account and allowance is made and for a 
possible outflow collimation and a lateral spread of the jet. When the blast-wave has decelerated
sufficiently that the jet edge is seen, the afterglow light-curve should exhibit an achromatic
break (steepening of the power-law decay). 

 Fox \etal (2005) claimed that the optical light-curve of the GRB afterglow 050709 exhibits a 
break at $t_b = 10$ d, though the post-break decay was not monitored to prove that a jet-break 
indeed existed. The fast decay of the radio light-curve of the GRB afterglow 050724 after 1 day 
suggests that a jet-break occurred at about $t_b = 2$ d, but
the possibility of a strong interstellar scintillation (GRB 050724 occurred at a low Galactic
latitude), the paucity of the radio measurements and the possible passage of the peak frequency
of the forward shock continuum through the radio domain at days after the burst prevent a
firm identification of the jet-break epoch.

 If the jet-break epoch is known, with the aid of the kinetic energy $E$ and circumburst density 
$n$ resulting from afterglow modelling, one can constrain the initial jet half-angle $\theta_{jet}$.
Starting from the expression of the shock's Lorentz ($\Gamma$) as a function of observer time -- 
$\Gamma = 6.5\, (E_{50}/n_0)^{1/8} t^{-3/8}$, where $E_{50}$ is the outflow's kinetic energy 
in $10^{50} \ergsr$, $n_0$ the medium density in protons per $\cm3$ and $t$ is the observer time 
(in days) corresponding to the arrival time of the photons emitted from the fluid moving at an 
angle $\Gamma^{-1}$ relative to the observer's line of sight toward the blast-wave center -- 
the condition that the jet edge becomes visible at some time $t_b$ (and that the effects of 
collimation are manifested): $\theta_{jet} = [\Gamma (t_b)]^{-1}$ leads to
\begin{equation}
 \theta_{jet} = 9\deg \, (n_0/E_{50})^{1/8} t_b^{3/8} \;.
\label{thjet}
\end{equation}
This implies a the jet energy 
\begin{equation} 
 E_{jet} = \pi \theta_{jet}^2 E \sim \times 10^{49} n_0^{1/4} (E_{50} t_b)^{3/4} \; {\rm erg} \;.
\label{Ejet}
\end{equation}

 The electrons of the shocked circumburst medium radiate synchrotron emission and upscatter it. 
The emitted flux depends on three free parameters pertaining to the magnetic field energy 
($\epsilon_B$), the minimum electron energy ($\epsilon_i$) and the power-law index of the 
electron distribution with energy ($p$) in the shocked medium. All are assumed to be constant. 
Together with the blast-wave energy and the medium density, the $\epsilon_B$ and $\epsilon_i$
parameters determine the peak flux of the synchrotron continuum ($F_p$) and its breaks: the 
self-absorption frequency ($\nu_a$), the synchrotron characteristic frequency for the minimum 
energy electrons ($\nu_i$) and the cooling frequency ($\nu_c$) at which emit the electrons whose 
radiative and adiabatic timescales are equal. The parameter $p$ determines the slope $\beta$ 
of the synchrotron continuum above its peak and the index $\alpha$ of the power-law decay of 
the afterglow flux ($F_\nu (t) \propto \nu^{-\beta} t^{-\alpha}$).
It follows that the relative intensities of the measured radio, optical and X-ray fluxes and
their decays constrain the five afterglow parameters: two ($E$ and $n$) for the dynamics of 
the blast-wave and three ($\epsilon_i$, $\epsilon_B$ and $p$) for the synchrotron radiation. 

 As can be seen from equation (\ref{thjet}), the epoch of the optical light-curve break 
constrains primarily the initial jet aperture. Because the shape and duration of the optical 
light-curve steepening also depends on how relativistic is the jet at the break time, a 
numerical fit to the light-curve steepening also provides constraints on the blast-wave energy 
and circumburst density. Using this afterglow model, Panaitescu (2005) has found the following
parameters for a sample of ten long-GRB afterglows whose optical light-curves exhibited a 
break at $t_b \sim 1$ d after the burst: $E = 10^{52}-10^{53} \ergsr$, $\theta_{jet} = 2\deg-3\deg$ 
and $E_{jet} = (2-6) \times 10^{50}$ erg.

\section{The GRB Afterglow 050709}
\label{050709}

\begin{figure}
\vspace*{5mm}
\psfig{figure=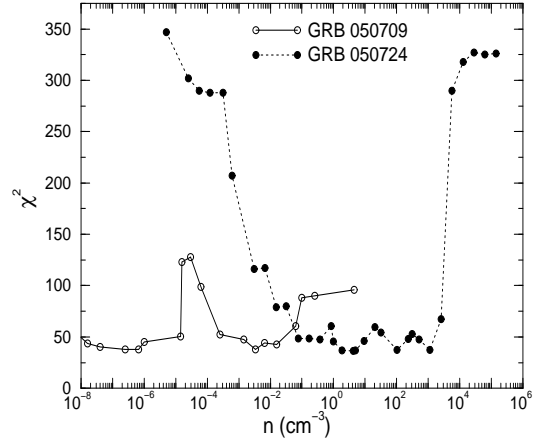,width=7cm,height=6cm}
\caption{ Variation of the $\chi^2$ of the best fit to the multiwavelength emission of 
    the GRB afterglows 070509 (17 measurements) and 050724 (24 measurements) with the 
    density ($n$) of the uniform medium. For a fixed $n$, the other four model parameters 
    (see \S\ref{aglow}) were left free to minimize the $\chi^2$ between model fluxes and 
    observations.  }
\label{nchi2}
\end{figure}

\begin{figure}
\vspace*{5mm}
\psfig{figure=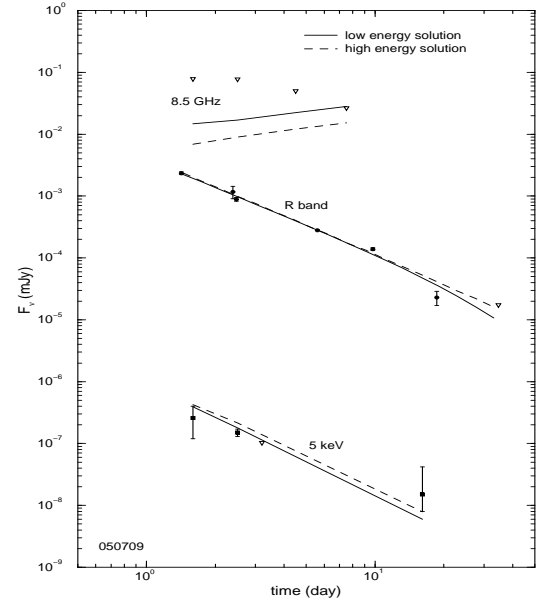,width=7cm,height=8cm}
\caption{ Fits to the optical and X-ray measurements of GRB afterglow 050709 ($z=0.16$), 
    satisfying the radio upper limits (2$\sigma$). Data are from Covino \etal (2005), 
    Fox \etal (2005) and Hjorth \etal (2005). 
    Two X-ray measurements at 2.8 d and 16 d, showing a flaring behaviour have been excluded,
    as it is likely that they are due to a different mechanism (internal shocks). 
    The {\sl solid} curves are for a fit with a lower blast-wave kinetic energy per solid 
    angle, $E = 3\times 10^{49} \ergsr$, a medium density $n = 6\times 10^{-3}\cm3$ and a 
    jet half-angle $\theta_{jet} > 12\deg$, as required by the lack of a jet-break until 
    10 days - eq. [\ref{thjet}]), corresponding to a jet energy $E_{jet} > 4\times 10^{48}$ 
    erg (eq. [\ref{Ejet}]).
    The {\sl dashed} curves show a fit for a higher blast-wave energy, $E = 3\times 10^{52} 
    \ergsr$, a very low medium density, $n = 2\times 10^{-7}\cm3$ and $\theta_{jet} > 1.4\deg$, 
    for which the jet energy is $E_{jet} > 6\times 10^{49}$ erg.  }
\label{0507}
\end{figure}

 The lack of radio detections and an optical light-curve break for this afterglow prevent us
from obtaining tight constraints on its physical parameters. As shown in Figure \ref{nchi2}, 
the best fit $\chi^2$ as a function of the circumburst medium density has two broad minima
for which $n < 10^{-5} \cm3$ and $2\times 10^{-4} \cm3 < n < 5\times 10^{-2} \cm3$. An example 
of a fit for each type of density is shown in Figure \ref{0507}.

 For the low density solutions, the outflow has a kinetic energy per solid angle $E = (3-300) 
\times 10^{50} \ergsr$, hence the 30--400 keV burst output ($E_\gamma = 5\times 10^{48} \ergsr$ 
-- Fox \etal 2005) implies a GRB efficiency below 1 percent. From equations (\ref{thjet}) 
and (\ref{Ejet}), the lack of a jet-break until 10 days requires that $\theta_{jet} > 1.5\deg$ 
and $E_{jet} > 10^{48}$ erg. The magnetic field energy ($\epsilon_B$) and the minimum electron 
energy ($\epsilon_i$) are found to be larger than $10^{-3}$ and 0.04 of the energy of the 
post-shock fluid, respectively. 

 We note that, for such low ambient densities and at the redshift $z=0.16$ of this burst, 
the apparent source size is $\theta (t) \sim 20\, (E_{52}/n_{-6})^{1/8} t^{5/8}\, {\rm \mu as}$ 
(with the notation $Q ({\rm cgs}) = 10^n Q_n$ and time measured in days). The GRB afterglow 
050709 was not detected in the radio down to $30\, {\rm \mu Jy}$. However the radio emission 
of other small-redshift afterglows interacting with so tenuous media may be bright enough 
to be detected (the peak radio flux is above 0.3 mJy for a burst redshift $z < 0.4\, E_{52}^{1/2} 
n_{-6}^{1/4} \epsilon_{B,-1}^{1/4}$), which would provide two tests for a low density circumburst 
medium. First, the source size at $t > 10$ days could be sufficiently large to measure with the 
VLBA. Second, above 1 GHz, the source is larger than the size of the first Fresnel zone for weak 
scattering by the Galactic interstellar medium and larger than the size of the strong scattering 
disk, consequently the amplitude of radio scintillations should be small. 

 For the higher density solutions, the resulting outflow kinetic energy density is 
$E = (2-10) \times 10^{49} \ergsr$, therefore the GRB efficiency is between 5 and 20 percent.
A jet-break after 10 days requires $\theta_{jet} > 6\deg$ and $E_{jet} > 3\times 10^{48}$ erg. 
The microphysical parameters have values slightly below equipartition. We note that these 
afterglow parameters are similar to those found by Fox \etal (2005).

 That the fit microphysical parameters increase with the medium density (while the outflow 
energy decreases) and approach equipartition for $n \sim 0.1\cm3$ shows why a higher external 
density is not allowed. Put in equivalent form, as the medium density is increased, the cooling
frequency falls below the optical, leading to an incompatibility between the model and
observational optical--to--X-ray flux ratios: for $\nu_c < \nu_o$, the 1--10 days power-law 
decay of the optical emission, $F_o \propto t^{-1.48 \pm 0.06}$, implies that the {\sl intrinsic} 
optical spectral energy distribution has a slope $\beta = (2\alpha_o+1)/3 = 1.32 \pm 0.04$ 
larger than the optical--to--X-ray spectral slope, $\beta_{ox} = 1.11 \pm 0.02$ (calculated from 
observations at 1.6 d, 2.5 d and 16 days), a situation which cannot be accommodated by the standard 
blast-wave scenario.

\section{The GRB Afterglow 050724}
\label{050724}

 The X-ray emission of the GRB afterglow 050724 exhibits a flare at 0.2--3 d after the burst 
(Burrows \etal 2005). Such flares, having a timescale comparable to the epoch when they occur, 
have been observed for many other bursts. The X-ray flares with the shortest duration, 
$\delta t/t_{on} \siml 1$ are unlikely to arise from the afterglow blast-wave, as all plausible 
mechanisms related to it (energy injection, hot spots, reverse shock emission) should produce 
either longer-lived or weaker flares. Instead such short-lived flares may be due to late internal 
shocks (Zhang \etal 2005) occurring in an outflow ejected after that which produced the burst. 
For this reason, we model first the afterglow 050724 emission without the X-ray flare. 

\begin{figure}
\vspace*{5mm}
\psfig{figure=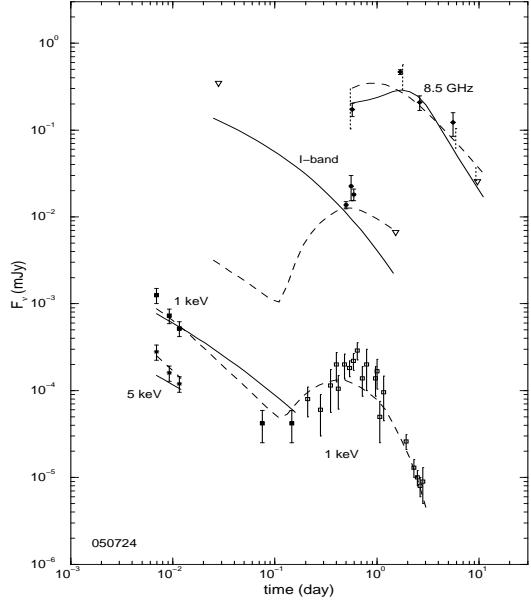,width=7cm,height=8cm}
\caption{ Fits to the radio, optical and X-ray measurements of GRB afterglow 050724 ($z=0.26$).
     Radio data were provided by the GRB Large Program at the VLA 
     (http://www.vla.nrao.edu/astro/prop/largeprop -- D. Frail). 
     Optical data are from Berger \etal (2005) and GCNs 3670 (M. Chester) and 3699 (K. Wiersema).
     Optical fluxes have been corrected for a Galactic reddening of $E(B-V) = 0.26$ (Gorosabel 
     \etal 2005). The X-ray fluxes were inferred from the 0.2--10 keV emission reported by
     Burrows \etal (2005) and the spectral slope of GCN 3669 (P. Romano). 
     The {\sl solid} curves show a fit to the data without the X-ray flare at 0.2--3 d,
     with parameters $E = 5\times 10^{49} \ergsr$ and $n = 5\cm3$, hence a jet-break at 
     $t_b = 1$ d implies $\theta_{jet} = 12\deg$ and $E_{jet} = 7 \times 10^{48}$ erg. 
     Vertical dotted lines indicate the amplitude of Galactic interstellar scintillation at 8.5 GHz. 
     The {\sl dashed} curves indicate a fit where $E = 4 \times 10^{50} \ergsr$ are injected 
     into the blast-wave at 0.1--0.4 d, to accommodate the 0.2--3 d X-ray flare (open symbols). 
     The medium density is unchanged, $\theta_{jet} = 10\deg$ and the jet energy after injection 
     is $E_{jet} = 4 \times 10^{49}$ erg.  } 
\label{0724}
\end{figure}

 The paucity of optical measurements for this afterglow prevents a tight constraining of its 
parameters. The medium density which can accommodate the emission of this afterglow is in the 
$10^{-1}\cm3 - 10^3 \cm3$ range (Figure \ref{nchi2}). The ejecta kinetic energy per solid angle 
is $E = (1-30) \times 10^{49}\ergsr$ thus, if the jet-break epoch is $t_b \sim 1$ d (as suggested 
by the steep $t^{-2}$ decay of the radio emission), then $\theta_{jet} = 10\deg-15\deg$ and 
$E_{jet}= 10^{48}-10^{50}$ erg. The burst output, $E_\gamma = 3\times 10^{49} \ergsr$ (Berger 
\etal 2005), implies that the GRB efficiency is between 10 and 75 percent. The magnetic and 
minimum electron energies are both found to be larger than 0.01 of the shocked medium energy. 
A fit with such parameters is shown with solid curves in Figure \ref{0724}. 

 The bounds on the allowed medium density arise from the condition that the magnetic and electron 
energies do not exceed equipartition. The equality of the optical spectral slope $\beta_o = 0.99 
\pm 0.08$ (inferred from the $K$-, $I$-, $R$- and $V$-band measurements at 0.5 day) to the 
optical--to--X-ray spectral slope, $\beta_{ox} = 0.89 \pm 0.04$ (calculated from the extrapolation 
of the 0.01--0.1 day X-ray fluxes at 0.5 day), implies that, at 0.5 day, the cooling frequency 
($\nu_c$) is not between the optical and X-ray domains. Consistency between the X-ray decay 
index ($\alpha_x = 0.8 \pm 0.2$) and the spectral slope ($\beta_x = 0.93 \pm 0.05$) reported
in GCN 3669 (P. Romano), requires that $\nu_c$ is below the X-rays, hence it must also be below 
the optical ($\nu_c < 10^{15}$ Hz). The rise and fall of the radio emission before and after 
2 d, respectively, indicate that the peak frequency of the forward-shock emission crosses the 
radio domain at that epoch, hence $\nu_i (t=2 {\rm d}) \sim 10$ GHz, and that the peak flux is 
$F_p \sim 0.5$ mJy (constant before the jet-break time $t_b$). From the expressions of the synchrotron 
continuum characteristics ($\nu_a$, $\nu_i$, $\nu_c$, $F_p$) as functions of the afterglow parameters 
($E$, $n$, $\epsilon_i$, $\epsilon_B$), the above constraints lead to $n \sim 1\, \nu_{a,10}^{4.2}
\nu_{c,15}^{0.7} \cm3$, $E \sim 10^{50} \nu_{a,10}^{-0.8} \nu_{c,15}^{0.2} \ergsr$, $\epsilon_i 
\sim 0.02\, \nu_{a,10}^{0.8} \nu_{c,15}^{0.2}$ and $\epsilon_B \sim 0.03\, \nu_{a,10}^{-2.5} 
\nu_{c,15}^{-1.2}$, where $\nu_a$ and $\nu_c$ are in $10^9$ and $10^{15}$ Hz, respectively. 
Then, the conditions $\epsilon_i < 0.1$ and $\epsilon_B < 0.1$ lead to $\nu_a \sim 10$ GHz 
(constant before $t_b$), with a weak dependence on $\nu_c$, and to that $n$ and $E$ are in the
above ranges found numerically through afterglow modelling. 


 Given that the X-ray flare of this afterglow is long-lived, \ie it could arise from the forward 
shock, we have also modelled it with an episode of energy injection in the blast-wave.
Such fits (an example is shown in Figure \ref{0724} with dashed lines) require a larger energy
per solid angle after injection, $E = (4-40)\times 10^{50} \ergsr$, and the same medium density 
range as for the fit without the X-ray flare, hence the jets is slightly narrower, $\theta_{jet} = 
8\deg-12\deg$, and the jet energy larger, $E_{jet} \simg (1-50) \times 10^{49}$ erg.

\section{Conclusions}

 The circumburst medium density ($n < 10^{-5} \cm3$) of the high-energy solution for the 
multiwavelength emission of the GRB afterglow 050709 is consistent with the hypothesis 
that this burst arose in a binary merger, as such mergers can occur well outside the host galaxy 
environment. Its radio emission has not been detected; a brighter radio afterglow at low-redshift 
($z < 0.4$) and occurring in the intergalactic medium could provide two direct tests for the 
existence of very tenuous media around short GRBs: the radio source would be sufficiently large 
to measure with the VLBA and to quench the interstellar scintillation. 

 However, modelling of the GRB afterglow 050709 emission does not provide a definitive argument 
in favour of the binary merger scenario, as the highest allowed density (obtained for the low-energy 
solution discussed in \S\ref{050709}), $n \siml 0.1 \cm3$, is comparable with that of the interstellar 
medium. Taking into account that this afterglow is at only 4 kpc (in projection) from the center of 
its host, the higher density solution appears more plausible.

 The medium density which accommodates the emission of the GRB afterglow 050724, $n = 10^{-1} -
10^3 \cm3$, is larger than for a galactic halo or the intergalactic medium and consistent 
with that this afterglow is at 2.5 kpc from the host's center. This result does not lend support 
to the binary merger origin for GRB 050724 but it does not rule it out either, as it is possible 
that a low binary velocity combined with a short merging time or a strong host galaxy gravitational 
potential led to a merger occurring in the intragalactic environment.

 The radio measurements of the afterglow 050724 indicate that the jet-break is at about 1 day
which, for the medium density and ejecta kinetic energy per solid angle ranges identified through
afterglow modelling, requires that the jet is wider than $8\deg$ (half-opening). The optical 
emission of the afterglow 050709 suggests that the jet-break is at or after 10 days which,
for an ejecta kinetic energy per solid angle comparable to that of the afterglow 050724
(\ie the low-energy/high-density solution discussed in \S\ref{050709}), requires a jet half-angle 
larger than $6\deg$. 

 Compared with the properties of the outflows of long-GRB afterglows inferred by Panaitescu 
(2005), the jets of the short GRBs 050709 and 050724 are at least twice wider, have a kinetic
energy per solid angle smaller by a factor of 100 or more and a collimation-corrected
kinetic energy which may be smaller by a factor of about 5--50. 

 If confirmed with other future afterglows, the weaker collimation of short-burst outflows 
will indicate that GRB jets are not collimated by magnetic fields but, we should 
keep in mind that the jet penetration through the He envelope could enhance significantly the 
collimation of the outflows produced in collapsars. The smaller kinetic energy of the short-burst
outflows would also suggest that GRB jets are powered by the gravitational binding energy of the 
accretion disk and not by the BH spin energy, \ie the disk's magnetic field does not thread the BH. 
This suggestion is based on that the expectation that debris disk resulting in NS-NS/BH-NS mergers 
is 10 times less massive than those of collapsars, while the BH spin energy should be larger
in the former model than in the latter.
 
 However, the existence of some underluminous long bursts (GRB 980425 -- Galama \etal 1998, 
GRB 020903 -- Sakamoto \etal 2004, 031203 -- Sazonov, Lutovinov \& Sunyaev 2004), whose output 
was at least 100 times smaller than that of the majority of long bursts 
casts some doubt on the existence of a dichotomy of the short and long bursts, based on their 
collimation and energetics. Future observations of short-GRB afterglows and a better coverage 
at radio frequencies and of light-curve jet-breaks, will allow us to extract through afterglow 
modelling tighter constraints on their jet properties and to make a more robust comparison with 
those of long bursts. The same approach will also determine the circumburst density, leading to 
an independent test for the origin of short bursts in compact binary mergers, provided that
these mergers occur in the intergalactic medium.


\end{document}